\documentclass[12pt]{iopart}
\usepackage{iopams}
\usepackage{graphicx}
\graphicspath{ {images/} }
\usepackage{epstopdf}
\usepackage{hyperref}

\usepackage{color}
\definecolor{green}{rgb}{0,0.5,0}
\usepackage{textcomp}

\def\beq{\begin{eqnarray}}
\def\eeq{\end{eqnarray}}
\def\be{\begin{equation}}
\def\ee{\end{equation}}
\def\eq{&=&}

\def\bm{\begin{math}}
\def\me{\end{math}}
\def\bi{\bibitem}

\def\bel{\begin{equation} \label}
\def\beel{\begin{eqnarray} \label}

\newcommand \nn {\nonumber}
\newcommand \bei {\begin{itemize}}
\newcommand \eei  {\end{itemize}}

\newcommand \nt   {\nonumber \\ }

\renewcommand{\leq}{\leqslant}
\renewcommand{\geq}{\geqslant}

\usepackage{amsthm}
\theoremstyle{definition}
\newtheorem*{definition}{Definition}

\begin{document}

\title{Exact solutions of temperature-dependent Smoluchowski equations} 

\author{A. I. Osinsky$^{1}$, N. V. Brilliantov$^{2}$}

\address{$^{1}$Marchuk Institute of Numerical Mathematics of Russian Academy of Sciences, Moscow, 119333 Russia}
\address{$^{2}$Department of Mathematics, University of Leicester, Leicester LE1 7RH, United Kingdom}

\begin{abstract}
We report a number of exact solutions for temperature-dependent Smoluchowski equations. These equations quantify the ballistic agglomeration, where the evolution of densities of agglomerates of different size is entangled with the evolution of the mean kinetic energy (partial temperatures) of such clusters. The obtained exact solutions may be used as a benchmark to assess the accuracy and computational efficiency of the numerical approaches, developed to solve the temperature-dependent Smoluchowski equations. Moreover, they may also illustrate the possible evolution regimes in these systems. The exact solutions have been obtained for a series of model rate coefficients, and we demonstrate that there may be an infinite number of such model coefficient which allow exact analysis. We compare our exact solutions with the numerical solutions for various evolution regimes; an excellent agreement between numerical and exact results proves the accuracy of the exploited numerical method. 
\end{abstract}


\submitto{\jpa}

\maketitle

\section{Introduction}  
Aggregation phenomena, when two objects of different size  meet each other and form a joint aggregate,  are widely spread in nature, e.g. \cite{BrilliantovPNAS2015, Erik17, BlumErosion2011,FalkovichNature,Falkovich2006,muller1928allgemeinen,Smo17,Igor14,Leyvraz2003,krapbook,BFP2018,Das,Mazza,BOK_PRE20,OB_PRE2022,BOK_PRL}.  The  spatial and time scales of such processes span many orders of magnitude. They occur in astrophysical systems, like galaxies clustering \cite{Galaxies, oort1946gas}, and in everyday-life, like  blood clotting \cite{bloodclott}, or  curdling of milk \cite{Colloid1,Colloid2}. Aggregation is also  ubiquitous on microscopic, molecular scales, see e.g. \cite{Erik17,Igor14,RW04,Prions2003}.  In dilute systems, where the objects collide mainly pairwize, the agglomeration kinetics is described by the infinite set of Smoluchowski equations \cite{smoluchowski-deu}. These equations quantify the evolution of the aggregates densities $n_k(t)$, where $k$ is the size of the aggregate, characterized by its mass, or equivalently by the number of monomers (the elementary objects) comprising the agglomerate. The  Smoluchowski equations \cite{Smo17,Smo16},  read: 

\begin{eqnarray}
\label{Smol}
\frac{\rmd n_k}{\rmd t} = \frac{1}{2}\sum_{i+j=k}C_{ij}n_in_j-n_k\sum_{j\geq 1}C_{kj}n_j. 
\end{eqnarray}
The rate coefficient $C_{ij}$ in the above equations give the reaction rates (number of reactions in unit volume per unit time) for the agglomeration process, $[i]+[j] \to [i+j]$.  The meaning of this equation is straightforward -- while the  first term in the right-hand side of (\ref{Smol}) quantifies  the increase of the density of aggregates of size   $k$ by the agglomeration, $[i]+[j] \to [k]$,  the second term quantifies the decrease of the density $n_k$, as all the reactions of such aggregates with other aggregates or monomers change the size $k$. Equations (\ref{Smol}) describe  space uniform systems; correspondingly, $n_k(t)$ are the average densities, that is, the density fluctuations are not included \cite{BOK_PRL}.

The rate coefficients $C_{ij}$ are determined by the transport mechanism which brings the aggregating particles together to the distances at which they can agglomerate. Basically, there are two main transport mechanisms -- diffusional and ballistic \cite{Leyvraz2003,krapbook}. The former is observed mainly in solutions, when aggregating particles, say solutes, collide many times with  solvent particles, before they undergo agglomerating collision with other solute particles. The latter mechanism is observed in moderately dense and dilute gases, including granular gases, when each collision between particles may be agglomerating, depending on the collision parameters. It has been recently intensively investigated, e.g.  \cite{BrilliantovPNAS2015,BFP2018,Das,Mazza,BOK_PRE20,CarnevalePameauYoung:1990,Leivraz1D,TrizacHansen:1995,Frachebourg1999,LF00,Trizac:2003,BrilliantovSpahn2006,Das1,Mazza1}. 

The rates $C_{i,j}$ for the diffusional transport depend on the size of reacting particles and their diffusion coefficients.  For the case of instantaneous reaction (diffusion-limited aggregation) of colliding spherical aggregates of radii $R_i$ and $R_j$ the  reaction rates read  \cite{Leyvraz2003,krapbook},
\bel{Cij}
C_{ij} = 4 \pi (R_i+R_j)(D_i+D_j), 
\ee
where $D_i =  \frac{k_BT}{6 \pi \eta R_i}$, is the diffusion coefficient of $i$th particle. Here $T$ and $\eta$ are respectively the solvent temperature and viscosity and  $k_B$ is the Boltzmann constant \cite{Leyvraz2003,krapbook}.  $R_i$ depends on the  size of the cluster $i$ as $R_i = r_1 i^{1/d}$, where $d$ is the dimension of the aggregate (which can be fractal) and $r_1$ is the length constant, characterising the monomer size; it may also include the packing fraction of monomers inside the cluster. 

Generally, aggregation with the diffusional   transport may have the rates, $C_{ij}$, different from those of Eq. (\ref{Cij}),   for instance, for  reaction-limited aggregation. Moreover, convenient phenomenological expressions for  $C_{ij}$ are often used. Still, Smoluchowski equations (\ref{Smol}) form a closed set for the  number densities $n_i(t)$, as long as the external parameters (e.g. temperature, $T$, viscosity  of the solvent, $\eta$, etc.) are fixed.

Exact solutions of (\ref{Smol}) for real microscopic expressions for the rate coefficients are presently lacking, hence  numerical solutions of such equations are  required. Still, there exists a number of model rate kernels $C_{ij}$, allowing analytical results, namely, the solutions are known for a constant, additive and multiplicative kernels:
\begin{eqnarray}\label{exSE1}
& n_k = \left(1+ \frac12 t \right)^{-2} \left( \frac{t}{t+2} \right)^{k-1} \qquad & C_{ij} = 1; \\
\label{exSEipj}
& n_k = \frac{\left( (1-\rme^{-t}) k \right)^{k-1} \rme^{-(1-\rme^{-t})k}}{k!} \rme^{-t}   \qquad & C_{ij} = i + j; \\
\label{exSEitj}
& n_k = \frac{(tk)^{k-1} \rme^{-tk}}{(k+1)!} \qquad & C_{ij}= ij.
\end{eqnarray}
The above solutions are given for mono-disperse initial conditions, $n_k(0)= \delta_{k,1}$.  A linear combination of these kernels, 
\bel{Cijex}
C_{ij} = A +B(i+j) + C(ij), 
\ee
with arbitrary constants, $A$, $B$ and  $C$ also yields   the exact results  \cite{Leyvraz2003,krapbook}. The solutions (\ref{exSE1}) and (\ref{exSEipj}) apply for $0\leq t < \infty $ \cite{Smo16,Golovin}, while (\ref{exSEitj}) for $0\leq t \leq 1$ \cite{McLeod}; later the solution has been also found for $1\leq t < \infty $ \cite{Kokholm}. Exact solutions are  also available for a special class of kernels -- addition kernels \cite{BK91}

The existence of exact solutions for some class of kernels is very important, since such solutions help to assess an accuracy and computational efficiency of different numerical schemes. Moreover, exact solutions can unambiguously illustrate a variety of possible evolution  regimes for the whole time interval, while numerical solutions are, in principle, limited in time. 

 The second transport  mechanism of aggregation -- the ballistic agglomeration (BA) is conceptually more complicated.  Here the aggregation rates depend on  particles sizes and velocities -- the larger the velocities, the higher the aggregation rates. When particles merge and form a single aggregate,  the kinetic energy of their relative motion ``vanishes'' -- it transforms into heat. Hence, the total kinetic energy of all agglomerates permanently decreases. The decaying kinetic energy implies the decrease of the particles velocities and the  agglomeration slowdown. This simple reasoning demonstrates an important interconnection between the aggregation kinetics and evolution of kinetic energy of the system  \cite{BFP2018,BOK_PRE20,OB_PRE2022}. 

The kinetic energy per particle (the total kinetic energy, divided by the total number of aggregates) may be called the  average kinetic temperature $T$. In the  course of  agglomeration, clusters of different size $k$ emerge. Correspondingly, the average kinetic energy of   aggregates of size $k$ (their total energy over their number) may be called the  partial  temperatures of such species, $T_k$. Generally, $T_k$ may differ for different $k$. As it follows from the above discussion,  the rate coefficients $C_{ij}$ depend on the kinetic energies of particles of size $i$ and $j$, that is, $C_{i,j}=C_{i,j}(T_i,T_j)$. 

If we knew the partial temperatures $T_i$ for all $i$, the Smoluchowski equations for the densities $n_i(t)$ (\ref{Smol}) would remain closed. Such an assumption for the average velocities of clusters of size $i$, namely, $v_i \sim \sqrt{T_i}\sim i^{-1/2}$  has been exploited in Refs. \cite{CarnevalePameauYoung:1990,Leivraz1D}. It was justified by the mean-field arguments for the momentum conservation at the collisions and random nature of the particles velocities. This yields  for the rate coefficients in 3D, $C_{ij} \sim (R_i+R_j)^2|v_i -v_j| \sim (i^{1/3} + j^{1/3})^2 | i^{-1/2} -j^{-1/2}|$. Unfortunately, such an estimate of $C_{ij}$ is too crude, as it does not describe a number of aggregation regimes observed in the BA, see e.g. \cite{BFP2018,BOK_PRE20,OB_PRE2022}. 

To obtain more adequate aggregation  rates  for the BA, one needs to derive them from a microscopic kinetic equation, namely from the Boltzmann equation, e.g. \cite{BrilliantovPoeschelOUP}. Boltzmann equation describes is this case the  evolution of the mass-velocity distribution function $f(m_k, {\bf v}_k, t)$ for different ``species'' of size $k$. It accounts for bouncing collisions (which may be dissipative) and aggregating collisions  \cite{BFP2018,BOK_PRE20}. Using a standard approach (e.g. \cite{BrilliantovPoeschelOUP}) one can obtain from the Boltzmann equation equations for the moments of the function $f(m_k, {\bf v}_k, t)$. Zero-order moments correspond to the number density of the species, $n_k(t)$, while the second-order moments are associated with the partial temperatures of the species, $T_k(t)$. The derivation, detailed in Ref.  \cite{BFP2018}, leads to a coupled set of equations. The first set of equations for densities, $n_k(t)$, corresponds to  conventional Smoluchowski equations; the rates $C_{ij}(T_i, T_j)$ there depend on the partial temperatures, $T_i$ and $T_j$. The second set of equations describes the evolution of temperatures $T_k(t)$ and supplement Smoluchowski equations. To be more precise  in our statement, we give below the following definition.

\begin{definition}
We call (discrete) \textsl{temperature-dependent Smoluchowski equations (TDSE)} an 
infinite system of differential equations of the form:
\begin{eqnarray}
\label{n-eq}
& \frac{\rmd}{{\rmd t}}{n_k} = \frac{1}{2}\!\!\sum\limits_{i + j = k} \!\!C_{ij} \!\left( T_i, T_j \right) n_in_j  - \sum\limits_{j = 1}^\infty \!  C_{kj} \left( T_k, T_j \right) \!n_kn_j,  \\
\label{nt-eq}
& \frac{\rmd}{\rmd t} \!\!\left( {n_k}{T_k} \right) \! = \!\frac{1}{2}\!\! \sum\limits_{i + j = k}\! \!\! B_{ij} \!\left( T_i, T_j \right)\!n_in_j  
\!- \!\sum\limits_{j = 1}^\infty  \!D_{kj} \!\left( T_k, T_j \!\right)  n_kn_j,  
\end{eqnarray}
where $i$, $j$ and $k$ are positive integers and the kernels $C_{ij} \left( T_i, T_j \right)$, $B_{ij} \left( T_i, T_j \right)$ and $D_{ij} \left( T_i, T_j \right)$ are some real functions of two positive integer variables ($i$ and $j$) and two non-negative real variables ($T_i$ and $T_j$). Additionally, due to the physical reasons discussed below, the kernels $C_{ij}$ and $B_{ij}$ must be non-negative and symmetric, that is,  $C_{ij} \left( T_i, T_j \right) = C_{ji} \left( T_j, T_i \right) \geqslant 0$ and $B_{ij} \left( T_i, T_j \right) = B_{ji} \left( T_j, T_i \right) \geqslant 0$.
\end{definition}

Here $n_i(t)$ have the physical meaning of the densities of size-$i$ clusters and $T_i$ are the partial temperatures (average kinetic energies) of size-$i$ clusters.  $T_i$ characterize  the ensembles of particles of size $i$, where clusters can have different velocities ${\bf v}_i$, with some velocity distribution $f({m_i,  \bf v}_i,t)$.  The mean square average of ${\bf v}_i$ determines the temperature $T_i= \frac12 m_i\left< v_i^2 \right>$, where  $m_i=m_1 i$ is the mass of an aggregate of size $i$ ($m_1$ is the monomer  mass) and the angular brackets denote the ensemble average.  The microscopic expressions for the rate coefficients, $C_{ij}$, $B_{ij}$ and $D_{ij}$, may be obtained from the first principles, starting from the Boltzmann equation \cite{BFP2018}.  For completeness, the expressions for these coefficients are presented in Appendix A for a rather general collision model. 

The structure of TDSE (\ref{n-eq})-(\ref{nt-eq}) is dictated by their physical nature, supported by the microscopic derivation. Hence, the physical meaning of the rate coefficients is obvious. $C_{ij}$ have the conventional meaning. The coefficients $B_{ij}$ describe the rate  at which the kinetic energy density (energy per unit volume) of clusters of size $k$ increases due to aggregation of clusters of size $i$ and $j$. Correspondingly, $D_{ij}$ describe the rate at which the kinetic energy density of clusters of size $k$ change due to collisions (both aggregative and bouncing) with all other clusters. Similarly as for standard Smoluchowski equations, the rates $C_{ij}$,  $B_{ij}$ and $D_{ij}$ may be chosen phenomenologically, provided they fulfil the above constrains, following from their physical meaning.

Generally, the TDSE may be solved only numerically \cite{BFP2018,BOK_PRE20,OB_PRE2022}. They demonstrate a very rich behavior, including the  astonishing regimes of increasing temperature \cite{BFP2018,BOK_PRE20}, density separation and other \cite{OB_PRE2022}. Still, it is very important to have exact solutions for the TDSE, say for  phenomenological  rate coefficients;  this is done in the present study. The exact solutions can reveal different evolution regimes and serve as a benchmark in assessing the  accuracy of numerical methods. 

 The rest of the paper is organized as follows. In the next Section II we consider the most simple case of temperature equipartition for different clusters size and size-independent rate coefficients. In Section III the general case of different temperatures for different species is addressed. We discuss the class of model kernels which allow exact solutions of the TDSE and present the solutions to these equations for some representative cases. Finally, in Section IV we summarize our findings. 

\section{Exact solution of temperature-dependent Smoluchowski equations for equal partial temperatures }

If ballistic agglomeration of particles occurs not at each collision, but  only in a small fraction of collisions, that is, with some probability $q\ll 1$, the mean kinetic energy of aggregates of different size equilibrate, and the condition $T_i(t)=T(t)$ holds true for all $i$ (note that the common temperature is time-dependent). In this case, (\ref{nt-eq}) reduce to a single equation for $T(t)$. Then the corresponding equations read, 
\beel{TDSE1}
\label{TDSE10}
\frac{\rmd}{\rmd t} n_k &=& \frac12 \sum_{i+j=k}C_{ij}(T)n_in_k - \sum_{j \geq 1} C_{kj}(T) n_kn_j \\
\label{TDSE11}
\frac{\rmd}{\rmd t} NT &=& - \sum_{i,j} P_{ij}(T) n_i n_j ,
\eeq
where $N(t)= \sum_{i\geq 1}n_i(t)$ is the total cluster density and $P_{ij}(T)$ is expressed in terms of the rate coefficients $B_{ij}(T)$ and $D_{ij}(T)$:
$$
P_{ij}=\frac12\left[D_{ij}+D_{ji} - B_{ij} \right], 
$$
 see \cite{BFP2018} for explicit expressions. Here we start from the physical condition of dominantly agglomerating  collisions, which is realized when aggregation energy is much larger than the  kinetic energy of the aggregates, see the Appendix A for detail.  The rate coefficients significantly simplify in this case :
\bel{CPij}
C_{ij} = 2 \sigma_{ij}^{d-1} \sqrt{2 \pi T/\mu_{ij}} ; \qquad \qquad P_{ij} =\frac23 T C_{ij}, 
\ee
where $d$ is the dimension (here we focus on $d=3$), $\sigma_{ij}=r_1(i^{1/d}+j^{1/d})$ is the  collision cross-section and $\mu_{ij}=m_im_j/(m_i+m_j)$ is the reduced mass of the colliding pair. 


\subsection{Existence of the solutions with equal temperatures} 

In the previous section we did not question the existence of solutions of the TDSE with equal partial temperatures, relying on the physical argument of small aggregation probability at collisions. However, this argument is not helpful if we search for the exact solutions of the original temperature-dependent Smoluchowski equations. 

Here we address this problem more formally. Namely, we ask ourselves, whether there are cases, when TDSE (\ref{n-eq})-(\ref{nt-eq}) with the rate kernels $C_{ij}(T_i, T_j)$,   $B_{ij}(T_i, T_j)$ and $D_{ij}(T_i, T_j)$ can be exactly replaced by the system (\ref{TDSE10})-(\ref{TDSE11}) for equal partial temperatures, $T_k=T$, for all $k$.

To answer this question, we  subtract (\ref{n-eq}),  multiplied by $T_k$, from (\ref{nt-eq}), which yields, 
\begin{eqnarray}\label{eq:eqcond0}
  n_k \frac{\rmd}{\rmd t} T_k & = \frac{1}{2} \sum\limits_{i+j = k} \left( B_{ij}(T_i,T_j) - T_{i+j} C_{ij} (T_i,T_j) \right) n_i n_j \\
  & - \sum\limits_{j = 1}^{\infty} \left( D_{kj} (T_k,T_j) - T_{k} C_{kj} (T_k,T_j) \right) n_k n_j. \nonumber
\end{eqnarray}
The latter equation may be further simplified if we imply a constraint on the kinetic coefficients of the form, $B_{ij} (T_i,T_j)=\frac12 ({T_i + T_j}) C_{ij}(T_i,T_j)$. As a result, the first term will be zero as long as $\frac12({T_i + T_j}) = T_{i+j}$, i.e., when we have equal temperatures.
%
%
%
Then (\ref{eq:eqcond0}) turns into the following condition:
\begin{equation}\label{eq:eqcond}
  \frac{\rmd}{\rmd t} T_k = - \sum\limits_{j = 1}^{\infty} \left( D_{kj} (T_k,T_j) - T_{k} C_{kj} (T_k,T_j) \right) n_j.
\end{equation}
Hence, we need to find the coefficients $C_{ij}$ and $D_{ij}$, such that the right-hand side of the above equation depends on $k$ only through some power of $T_k$, eventually yielding $\frac{\rmd}{\rmd t} T_k^{\gamma} = \frac{\rmd}{\rmd t} T^{\gamma}$ for some constant $\gamma$. And if $\frac{\rmd}{\rmd t} T_k^{\gamma} = \frac{\rmd}{\rmd t} T^{\gamma}$ for equal temperatures, then the temperatures always remain equal, because that means the right hand side of the differential equation involving $\frac{\rmd}{\rmd t} \left( T_i^{\gamma} - T_j^{\gamma} \right)$ is $0$, when $T_i = T_j$, so the difference $T_i^{\gamma} - T_j^{\gamma}$ can't become non-zero. 

For instance, the set of coefficients $C_{ij}=C_0 \left(\frac{1}{2} T_i + \frac{1}{2} T_j \right)^{\alpha}$, $B_{ij}= C_0 \left(\frac{1}{2} T_i + \frac{1}{2} T_j \right)^{\alpha+1}$ and $D_{ij}= C_0 \left(\frac{1}{2} T_i + \frac{1}{2} T_j \right)^{\alpha+1} + C_1 T_i^{\beta} $ with some exponents $\alpha$ and $\beta$ and constants $C_0$ and $C_1$ will comply with the condition (\ref{eq:eqcond}) and thus provide equal temperatures. When the temperatures are initially equal, these kernels reduce the TDSE to the Smoluchowski equations with constant coefficients. Although, one still should be careful in selecting the parameters, so that the system would not only have a solution, but also guarantee that the solution can be expressed analytically. In particular, the choice $\alpha=1/2$, $\beta = 3/2$ and $C_1 = C_0 / 6$ yields the size-independent kernels $C_{ij}$ and $P_{ij}$ in (\ref{TDSE2})-(\ref{TDSE2a}) from the next Section \ref{sec:ind}.

Another example of such kernels reads, 
\beel{addexample} C_{ij} \!\!\eq \!\!\left( i + j \right) \left( \frac12 T_i + \frac12 T_j \right)^{\alpha}, \nt 
B_{ij} \!\! \eq \!\! \left( i + j \right) \left( \frac12 T_i + \frac12  T_j \right)^{\alpha + 1}, \\ 
D_{ij}  \!\! \eq \!\! \left( i + j \right) T_i \left(\frac12  T_i + \frac12 T_j \right)^{\alpha} \!+\! j \left( T_i^{\beta} - T_i \right) \left( \frac12 T_i + \frac12 T_j\right)^{\alpha}\!. \nonumber   
\eeq 
They  reduce the TDSE to the Smoluchowski equations with $T_k=T$ and size-additive coefficients, which we will look at in Section \ref{sec:lin}.

The problem of finding the appropriate rate kernels to reduce the TDSE to the Smoluchowski equations with $T_k=T$ and size-multiplicative coefficients, is however more severe.  Although  we have analyzed the reduced model kernels for $C_{ij}$ and $P_{ij}$ and found the exact solutions in Section \ref{sec:mul}, we did not succeed  to find the expressions for the original kernels, $C_{ij}$, $B_{ij}$ and $D_{ij}$ which  comply with the condition  (\ref{eq:eqcond}).  Hence, we cannot exclude that the size-multiplicative rate kernels with temperature equipartition do not possess physically relevant solutions.

\subsection{Size-independent rate kernels}\label{sec:ind}
If we neglect the size dependence of the rate coefficients (\ref{CPij}), adopting the model kernels  
\bel{CPconstT}
C_{ij}=C_0T^{1/2}; \qquad \qquad  P_{ij}=\frac23 T^{3/2}C_0,
\ee
Equations (\ref{TDSE1}), (\ref{TDSE11}) take the form, 
\beel{TDSE2}
\frac{\rmd}{\rmd t} n_k &=&  T^{1/2} C_0 \left[ \frac12 \sum_{i+j=k}n_in_k - \sum_{i \geq 1}  n_kn_i \right]\\
\label{TDSE2a}
\frac{\rmd}{\rmd t} NT &=& - \frac23 T^{3/2} C_0 \sum_{i,j} n_i n_j = -\frac23T^{3/2}C_0N^2 .
\eeq
A  physical example of the size independent kernel (\ref{CPconstT})  corresponds to the BA for $d=2$. In this case the rates (\ref{CPij}) read,  $C_{ij}=2\sigma^{d-1}\sqrt{T/\mu_{ij}} \sim (i^{1/d}+j^{1/d})^{d-1} (1/i+1/j)^{1/2}T^{1/2}$.  The homogeneity exponent here is zero, $\lambda=(d-2)/d= 0$, that is, $C_{ai,aj}=a^{\lambda} C_{ij}=C_{ij}$. Hence the size dependence of $C_{ij}$ is suppressed and may be well approximated by a constant, $C_{ij}\simeq C_0 \sqrt{T}$.

Introducing now the new time variable,
\begin{equation}\label{tau}
  \tau  = C_0\int_0^t \sqrt{T(t^{\prime})} dt^{\prime}, 
\end{equation} 
such that 
\begin{equation}
\label{taudt} 
  \frac{\rmd \tau}{\rmd t}  = C_0 \sqrt{T}, \qquad \qquad \tau \left( t = 0 \right) = 0, 
\end{equation}
we recast the last equations into the form, 
\beel{ex1}
\frac{\rmd n_k}{\rmd \tau} &=& \frac12 \sum_{i+j=k} n_in_j - \sum_{i \geq 1} n_k n_i \\
\label{ex2}
\frac{\rmd NT}{\rmd \tau}  \eq -\frac23 TN^2,  
\eeq
Equation (\ref{ex1}) is the standard Smoluchowski equation  written for the time variable $\tau$. For the mono-disperse initial conditions, $n_k(0)= n_0\delta_{k,1}$ the solution reads, see (\ref{exSE1}) and  \cite{Leyvraz2003}, 
\bel{ex3}
n_k(\tau) = \frac{4 n_0 \left(n_0 \tau \right)^{k-1}}{(2+ n_0 \tau)^{k+1}}.
\ee 
Summing up (\ref{ex1}), we obtain, 
\begin{equation}
\frac{\rmd N}{\rmd \tau} = - \frac12 N^2, \nonumber
\end{equation}
which yields $N(\tau)= 2n_0/(2+n_0\tau)$.  Substituting this  into (\ref{ex2}) and solving, we find 
\bel{T}
T(\tau)= T_0/(1+n_0 \tau/2)^{1/3},
\ee
where  $T_0$ is the  initial temperature. Then $\tau(t)$ follows from substituting (\ref{T}) into (\ref{taudt}) and integrating it with the given initial condition:
\bel{taut}
\tau= 2(1+t/\tau_0)^{6/7}/n_0-2/n_0, 
\ee
where $\tau_0^{-1}=(7/12)C_0 n_0 \sqrt{T_0}$. Substituting $\tau(t) $ from (\ref{taut}) into (\ref{ex3}) we obtain the density dependence in laboratory time. Similarly, using $\tau(t)$ in Eq. (\ref{T}), we find the time dependence of temperature and, in the same way, for the total cluster density:   

\bel{ex4}
T(t) = \frac{T_0}{(1+ t/\tau_0)^{2/7}}, \quad \qquad N(t) = \frac{n_0}{(1+ t/\tau_0)^{6/7}}. 
\ee
For large $t \to \infty$ we have from the above equations, $T\sim t^{-2/7} \sim t^{-0.2857}$ and $N\sim t^{-6/7} \sim t^{0.8571}$. The above exponents are very close to the corresponding exponents $0.28$  and $0.85/0.86$ found in the MD/MC simulation of the reaction-limited BA in d=2  \cite{Trizac:2003}.

Comparing (\ref{exSE1}) and (\ref{ex3}), (\ref{taut}), we observe that while Smoluchowski equations predict $n_k \sim t^{-2}$ for $t \to \infty$, TDSE show a significantly slower density  decay in this limit, $n_k \sim t^{-12/7}$.  This follows from the decreasing aggregation rates with the decaying temperature, see Eq. (\ref{ex4}).

\begin{figure}
    \centering
    \includegraphics{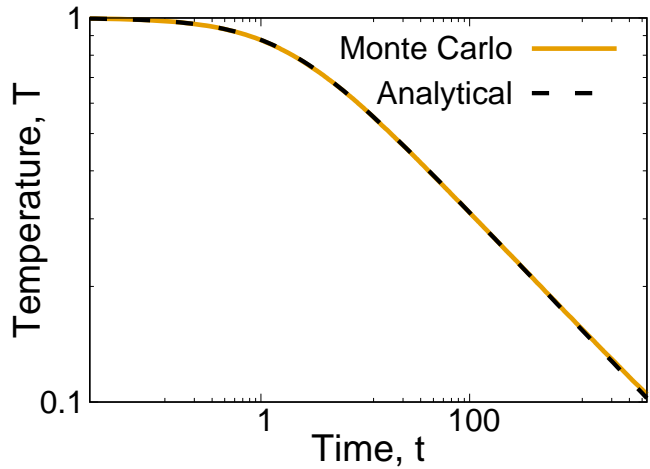}
    \caption{Comparison of the analytical solution, (\ref{ex4}), for the  temperature evolution $T(t)$ in the TDSE (\ref{TDSE2})-(\ref{TDSE2a}), with kernels (\ref{CPconstT}), with the results of Monte Carlo simulations obtained by method from \cite{OB_PRE2022} with $10^4$ particles.   Here $C_0 = 1$, $n_0 = 1$ and $T_0 = 1$.     }
    \label{fig:dec}
\end{figure}

To check the solution numerically, we use temperature-dependent Monte Carlo method \cite{OB_PRE2022}. Essentially, this method  is a straightforward extension of the standard Monte Carlo approach for the solution of Smoluchowski equations, e.g.  \cite{MCSmol,Babovsky,MCSmol2}. The evolution of temperatures in temperature-dependent Monte Carlo is treated in a similar way as the evolution of densities. It is computationally efficient and rather accurate approach \cite{OB_PRE2022}. We sketch the main ideas of this method in the Appendix B. The results for the temperature evolution are compared in Fig. \ref{fig:dec} with the exact solution (\ref{ex4}). A high accuracy of the temperature-dependent Monte Carlo approach of \cite{OB_PRE2022} is clearly visible.

\subsection{Size-additive rate kernels}\label{sec:lin}
Let us now abandon the constraint of high energy barrier and consider model rate kernels, still under the assumption of equal partial temperatures, $T_i=T$ for all $i$. We analyze the reaction rates,  additive with respect to the aggregates size, 
\bel{CBDipj}
C_{ij}= (i+j) T^{\alpha}; \qquad \qquad P_{ij}=\frac12(i+j) T^{\alpha +\beta},
\ee
with the positive $\alpha$ and $\beta$. Here for simplicity we omit dimension constants in front of $C_{ij}$ and $P_{ij}$; these may be put to unity by the corresponding rescaling of time and density. 

 Introducing, as in (\ref{tau}),  the new time variable, $\tau=\int_0^t T^{\alpha}(t')dt'$, we recast the system (\ref{TDSE1})-(\ref{TDSE11}) with the coefficients (\ref{CBDipj}) into the form: 
\beel{TDSE3}
\frac{\rmd n_k}{\rmd \tau} &=& \frac12 \sum_{i+j=k} (i+j) n_in_j - \sum_{j \geq 1} (k+j) n_k n_j \\
\label{TDSE31}
\frac{\rmd NT}{\rmd \tau}  \eq -T^{\beta}N.   
\eeq
Equation (\ref{TDSE3}) is the Smoluchowski equation for the new time variable $\tau$, for the additive kernel, with the solution, for densities, (\ref{exSEipj}) and total density \cite{Leyvraz2003}: 
\beel{nktau1}
  n_k \left( \tau \right) \!&=& \!\frac{k^{k-1}}{k!}\! \rme^{-\tau} \!\left(1 - \rme^{-\tau}\right)^{k-1} \!\exp \left(-k \!\left( 1 - \rme^{-\tau} \right) \right) \\
  \label{Ntau1}
  N(\tau) \eq \rme^{-\tau}, 
\eeq 
where again the mono-disperse initial condition $n_k(0)=\delta_{k,1}$ has been used; in what follows we will always use this initial condition, unless the opposite is indicated. Note that (\ref{Ntau1}) immediately follows from $\rmd N/ \rmd \tau =-N$, which results from (\ref{TDSE3}). Hence, (\ref{TDSE31}) takes the form, 
\bel{NTb}
\frac{\rmd NT}{\rmd \tau}=-(NT)^{\beta} \rme^{-(1-\beta) \tau},
\ee
with the solution (for $\beta \neq 1$), 
\bel{Tb}
T(\tau)= \left[ 1+ \left(T_0^{1-\beta}-1\right) \rme^{(1-\beta) \tau} \right]^{1/(1-\beta)}. 
\ee
Interestingly, depending on $T_0=T(0)$ and $\beta$ different evolution scenarios are possible. For the simplest one, $T_0=1$, temperature of the system keeps constant, $T(t)=1$ and the system evolves as for the common Smoluchowski equation for the additive kernel, as $\tau=t$. 

\begin{figure}
    \centering
    \includegraphics{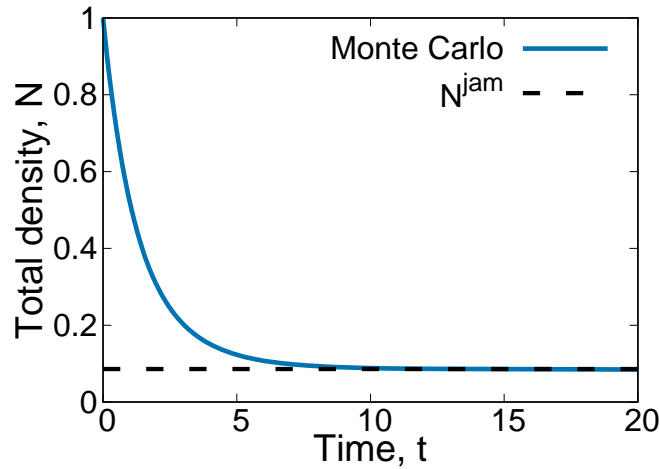}
    \caption{The solution for the total number density in TDSE (\ref{TDSE2})-(\ref{TDSE2a}), with kernels (\ref{CBDipj}), for $\alpha = \beta = 1/2$ and  $T_0 = 1/2$, obtained by the Monte Carlo method from \cite{OB_PRE2022} with $10^5$ particles. The convergence to the analytical result (\ref{jamN}) for the jammed total number density $N^{\rm jam} = 3/2 - \sqrt{2}$ is shown. }
    \label{fig:jam}
\end{figure}

(i) For $T_0<1$ and $\beta <1$, the system evolves until $\tau < \tau_*$, where $\tau_*= -\log (1-T_0^{1-\beta})/(1-\beta)$. At the modified time $\tau=\tau_*$ the temperature of the system turns to zero, along with the rate coefficients $C_{ij}$. At this moment the systems arrives to a jammed frozen state and its evolution ceases. The densities in the jammed state read: 
\beel{jam}
n_k^{\rm jam} \eq \frac{k^{k-1}}{k!} a \left(1 - a\right)^{k-1} \rme^{ -k \left( 1 - a \right) }; \\
\label{jamN}
N^{\rm jam} \eq  a =[1-T_0^{1-\beta} ]^{1/(1-\beta)}.
\eeq 
The laboratory time is related to the new time for $\tau < \tau_*$ as 
\bel{ttau}
t= \int_0^{\tau} \frac{\rmd \tau'}{T^{\alpha}(\tau')}=\int_0^{\tau} \frac{d u}{\left[ 1+b \rme^{(1-\beta)u} \right]^{\alpha/(1-\beta)}}, 
\ee
where $b= T_0^{1-\beta}-1$. Note that $b<0$ for $T_0<1$ and $\beta <1$. As $\tau \to \tau_*$, the laboratory time tends to infinity, $t \to \infty$, that is, the above jammed state (\ref{jam}) is approached asymptotically. The convergence of the number density to the final jammed state, as predicted above,  is demonstrated in figure \ref{fig:jam}. 

(ii) For $T_0>1$ and $\beta <1$, (\ref{Tb}) predicts an infinite increase of temperature as $\tau$ varies from $0$ to $\infty$. The laboratory time $t_*$, corresponding to $\tau=\infty$, is, however, finite: 
\bel{tinf}
t_*= \int_0^{\infty} \frac{\rmd u}{\left[ 1+b \rme^{(1-\beta)u} \right]^{\frac{\alpha}{1 - \beta}}}
\approx \frac{1}{b^{\frac{\alpha}{1 - \beta}} \alpha} - \frac{1}{b^{\frac{\alpha}{1 - \beta} - 1} (1-\beta)}, 
\ee
which implies an infinite increase of temperature, during a finite time interval, till the final time $t_*$. Somehow this effect resembles gelation -- a formation of an infinite cluster in a finite time, see e.g \cite{krapbook,Leyvraz2003}. It is not clear, however, whether this evolution scenario can correspond to any realistic physical process. 

\begin{figure}
    \centering
    \includegraphics{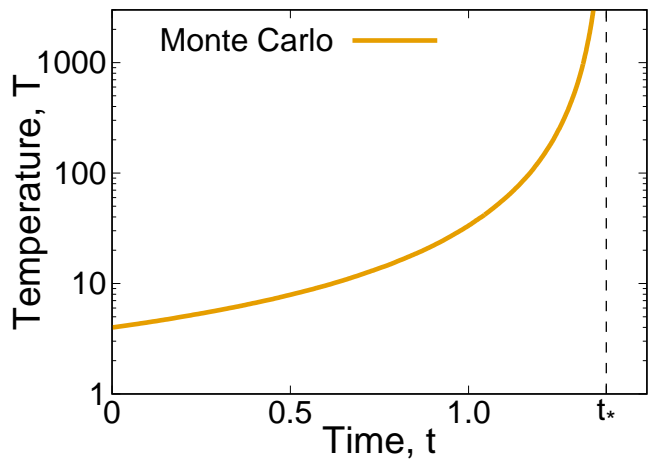}
    \caption{The solution for the temperature evolution in TDSE, (\ref{TDSE2})-(\ref{TDSE2a}), with kernels (\ref{CBDipj}), for $\alpha = \beta = 1/2$ and  $T_0 = 4$, obtained by the Monte Carlo method from \cite{OB_PRE2022} with $10^4$ particles. In agreement with the analytical predictions, (\ref{tinf}), temperature diverges  as  $t \to t_* = \ln 4$.}
    \label{fig:inf}
\end{figure}

The numerical solution for the temperature evolution  in this case is plotted in figure \ref{fig:inf} for  $\alpha = \beta = 1/2$ and $T_0 = 4$. As it is clearly seen from the figure, temperature indeed diverges within a finite time interval, in agreement witht the analytical prediction for $t_* = \int_0^{\infty} \frac{\rmd u}{1 + \rme^{u/2}} = \ln 4$.

\begin{figure}
    \centering
    \includegraphics{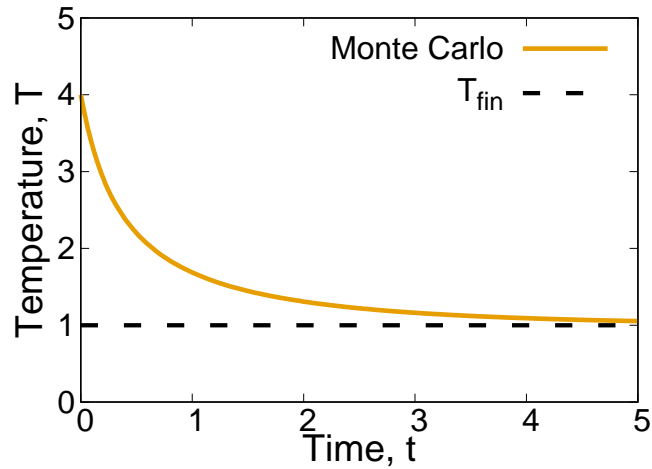}
    \caption{The solution for the temperature evolution in TDSE, (\ref{TDSE2})-(\ref{TDSE2a}), with kernels (\ref{CBDipj}) for  $\alpha = 1/2$, $\beta = 3/2$ and $T_0 = 4$, obtained by the Monte Carlo method from \cite{OB_PRE2022} with $10^4$ particles. In agreement with the analytical predictions temperature converges to the constant final temperature,  $T_{\rm fin} = 1$.}
    \label{fig:drop}
\end{figure}

(iii) For $T_0>1$ and $\beta >1 $, (\ref{Tb}) predicts the decrease of temperature from initial $T_0$ to the final temperature $T_{\rm fin}=1$. The relation between $t$ and $\tau$ is given by (\ref{ttau}), which may be written in terms of special functions. After a short relaxation time of the order of $\tau \sim 1/(1-\beta)$ we have $T\simeq 1$ and $\tau = t -{\rm const}$, with the densities given by (\ref{nktau1}). 
Temperature evolution for $\alpha = 1/2$, $\beta = 3/2$ and $T_0 = 4$ is plotted in figure \ref{fig:drop}; here $T$ converges to the constant final value of $T_{\rm fin} = 1$. 

(iv) For $T_0<1$ and $\beta >1 $, one has very similar behavior  to the above case (iii), with the only difference, that the temperature  initially increases, converging to the constant value of $T=1$.

\subsection{Size-multiplicative rate kernels}\label{sec:mul}
Consider now the rate kernels of the form, 
\bel{mul}
C_{ij}= (ij) T^{\alpha}; \qquad \qquad P_{ij}= (ij) T^{\alpha+\beta}.
\ee
and the respective rate equations, 
\beel{TDSE4}
\frac{\rmd n_k}{\rmd \tau} &=& \frac12 \sum_{i+j=k} (ij) n_in_j - \sum_{j \geq 1} (kj) n_k n_j \\
\label{TDSE41}
\frac{\rmd NT}{\rmd \tau}  \eq -T^{\beta}.   
\eeq
Here, the same definition of $\tau$ in the previous section is used. Equations (\ref{TDSE4}), (\ref{TDSE41}) are valid for $\tau <1$, as $\tau=1$ is the gelation point of the common Smoluchowski equations \cite{krapbook,Leyvraz2003}. The pre-gelling solution reads:
\beel{nkij}
n_k \eq \frac{(\tau k)^{k-1} \rme^{-\tau k}}{(k+1)!} \\
\label{Nij}
N(\tau) \eq  \left( 1- \frac12 \tau \right),  
\eeq
where again (\ref{Nij}) follows from the solution of $\frac{\rmd N}{ \rmd \tau}= - \frac12 $, resulting from (\ref{TDSE4}). From (\ref{TDSE41}) and (\ref{Nij}) we find, 
\bel{Tij}
T(\tau) = \left[2 + B_0 (1-\tau/2)^{-\beta_1} \right]^{1/\beta_1},
\ee
where we abbreviate, $\beta_1= 1-\beta$ and $B_0= T_0^{\beta_1}- 2$. Depending on $T_0$ and $\beta$ the following evolution scenarios may happen: 

(i) For $T_0>2^{1/\beta_1}$ and $\beta <1$ (\ref{Tij}) predicts temperature growth to $T(\tau=1)=2 \left[ B_0+2^{\beta} \right]^{1/\beta_1}$, associated with the gelation, which occurs at laboratory time, 
\begin{equation}
t_g=\int_0^1 \left[ 2+B_0 (1-\frac 12 u )^{-\beta_1} \right]^{-\alpha/\beta_1} du . \nonumber
\end{equation}
Note, however, that for the ballistic agglomeration with a true microscopic kernel the gelation is questionable, see \cite{OB_PRE2022}.

(ii) For $T_0<2^{1/\beta_1}$ and $\beta <1$ we have two different scenarios. If $T_0< \left(2 - 2^{\beta} \right)^{1/\beta_1}$, the temperature $T$ evolves to a jammed state at the modified time $\tau_*= 2-2^{-\beta/\beta_1} \left( 2 - T_0^{\beta_1} \right)^{1/\beta_1}<1$, where the temperature of the system is zero, $T(\tau_*)=0$. The jammed state occurs at the laboratory time
\begin{equation}
t_*= \int_0^{\tau_*} \left[ 2+B_0 (1- \frac12 u )^{-\beta_1} \right]^{-\alpha/\beta_1} du . \nonumber
\end{equation}
The density distribution of the jammed state is given by $n_k(\tau_*)$ with $n_k(\tau)$ defined in (\ref{nkij}). Another scenario for  $T_0>2^{\beta_1}-1 $ is similar to the one in the above item (i), except the temperature now decreases down to $T(\tau=1)=2 \left[ B_0+2^{\beta} \right]^{1/\beta_1}$. 

(iii) For $\beta >1$ gelation always happens, and we have qualitatively the same scenarios as in the items (i) and (ii). For $T_0>2^{1/\beta_1}$ we have gelation with cooling, for $T_0<2^{1/\beta_1}$ we have gelation with temperature increase.

%
%

\section{Exact solution of temperature-dependent Smoluchowski equations for different partial temperatures} 
Generally, temperatures of clusters of different size differ from each other and a complete set of equations for $T_i$ is needed. As has been mentioned above, it is worth to have exact solution for the complete TDSE, even for model coefficients, without  microscopic justification. This may be done for specially tailored rate kernels. An infinite number of such model kernels, which allow exact solution of TDSE may be found. All of them are based on the exact solutions of conventional Smoluchowski equation. We start with a certain example, when the TDSE may be  reduced to the Smoluchowski equations with size-independent kernels. Let the coefficients: 
\beel{kerex}
C_{ij} \eq \frac{T_i}{i} +\frac{T_j}{j}; \qquad B_{ij} = \left( \frac{T_i}{i} +\frac{T_j}{j} \right) (T_i+T_j) \nt 
D_{ij} \eq \left( 2\frac{T_i}{i} +\frac{T_j}{j} \right) T_i , 
\eeq
be the according rates for the full system
\begin{eqnarray}\label{eq:gentempsys}
  \frac{\rmd}{\rmd t} n_k & = \frac12 \sum\limits_{i + j = k} C_{ij}  n_i n_j - \sum\limits_{j \geq 1} C_{kj}  n_k n_j. \\
  \frac{\rmd}{\rmd t} \left(n_k T_k \right) & = \frac12 \sum\limits_{i + j = k} B_{ij}  n_i n_j - \sum\limits_{j \geq 1} D_{kj} n_k n_j. \nonumber
\end{eqnarray}
After multiplying the first set of equations by $T_k$, subtracting it from the second one and substituting the kernels (\ref{kerex}), we obtain, 
\beel{TDSEnkTk}
  n_k \frac{dT_k}{dt}  \eq \frac12 \sum\limits_{i + j = k} \left( \frac{T_i}{i} +\frac{T_j}{j} \right) \left( T_i + T_j - T_k \right) n_i n_j \nt 
  &-&  \sum\limits_{j \geq 1} T_k \left( \frac{T_k}{k} \right) n_k n_j.
\eeq
Let us search for the solution in the form $T_k(t) = k f(t)$. Then the first sum in the right-hand side of the above equation vanishes, and we arrive at 
\begin{equation}\label{eq:tf}
  \frac{\rmd}{\rmd t} f(t) = - f^2(t) \sum\limits_{j = 1}^{\infty} n_j = - f^2 N.
\end{equation}
With  this substitution, we also recast the first equation for $n_k$ in (\ref{eq:gentempsys})  into the form of standard Smoluchowski equation with the constant kernel, $C_{ij}=2$:
\[
  \frac{\rmd n_k}{\rmd \tau} = \frac12 \sum_{i+j=k}  2n_i n_j - \sum_{j \geq 1}  2n_k n_j,
\]
where we again introduce the new time variable,  
\be
  \tau = \int\limits_0^t f(t) dt. \nn
\ee
For the linear kernel the  exact solution  for mono-disperse initial condition is known, see (\ref{exSE1}), hence we write, 
\[
  n_k  \left( \tau \right) = \left(1+ \tau  \right)^{-2} \left( \frac{\tau }{\tau +1} \right)^{k-1}, 
\]
and $N(\tau) = (1+\tau)^{-1}$. Using the new time variable in (\ref{eq:tf}) we obtain,
\[
  \frac{\rmd}{\rmd \tau} f(\tau) = - f N.
\]
Then, substituting in the above equation $N(\tau) (1+\tau)^{-1}$ yields, for $N(0) = 1$ and $T_1(0) = f(0) = 1$, the solution
\[
  T_k(\tau) = \frac{k}{1 + \tau}, \qquad \tau = \sqrt{1+2t} - 1.
\]
As the result the densities for the above rate coefficients depend on the laboratory time as,  
\bel{nTex}
n_k(t) = \left( 1 - \frac{1}{\sqrt{1 + 2t}} \right)^{k-1} \left(1 + 2t\right)^{-1}.
\ee
As expected, TDSE predict significantly slower decay with time of the densities,  than Smoluchowski equations. This follows from the fact that in the course of time the motion of particles slows down as their temperatures decrease. 

\begin{figure}
    \centering
    \includegraphics{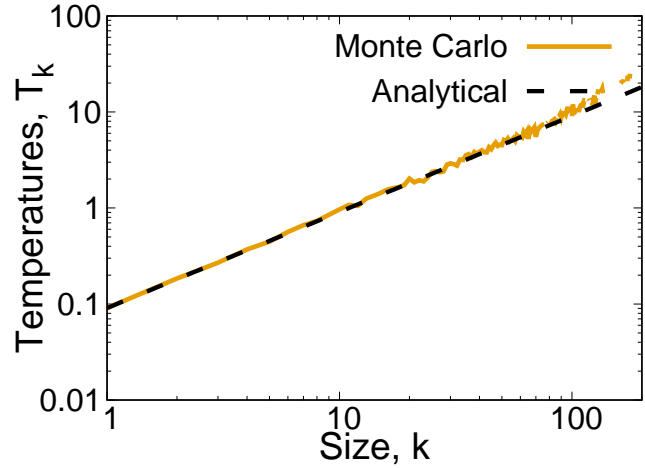}
    \caption{The distribution of  temperatures of  aggregates of size $k$, at $t = 10$. The solution of the TDSE (\ref{eq:gentempsys}) with kernels (\ref{kerex2}), obtained by the Monte Carlo method from \cite{OB_PRE2022} with $10^4$ particles, is compared with the  exact solution (\ref{eq:tempdist}).}
    \label{fig:tempdist}
\end{figure}

Other examples of kernels, which allow an exact solution for TDSE, read, 
\bel{kerex2}
C_{ij} = T_i +T_j; \quad B_{ij} = (T_i +T_j)^2; \quad D_{ij}= (T_i+T_j +1) T_i, 
\ee
and lead instead of (\ref{TDSEnkTk}) to the following equations, 

\bel{TDSEnkTk1}
  n_k \frac{dT_k}{dt}  = \frac12 \!\sum\limits_{i + j = k}\!\! \left(  T_i \!+ \!T_j \right) \left( T_i \!+ \!T_j\! - \!T_k \right) n_i n_j - \sum\limits_{j \geq 1} T_k n_k n_j, 
\ee
which with the same Ansatz for $T_k$ and the according modified time variable  $\tau$ lead, eventually,  to the common Smoluchowski equations with the additive kernel, $C_{ij}=i+j$:
\bel{exSEipj1}
  \frac{\rmd n_k}{\rmd \tau} = \frac12 \sum_{i+j=k} \left( i + j \right) n_i n_j - \sum_{j \geq 1} \left( k + j \right) n_k n_j,
\ee
with the solution,  
\[
  n_k \left( \tau \right) = \frac{k^{k-1}}{k!} \rme^{-\tau} \left(1 - \rme^{-\tau}\right)^{k-1} \exp \left(-k \left( 1 - \rme^{-\tau} \right) \right).
\]
Using the same steps as above and applying the initial conditions $n_k(0)=\delta_{k,1}$ and $T(0)=1$, we arrive for the solution for partial temperatures and densities in the laboratory time, 
\begin{equation}\label{eq:tempdist}
  T_k(t) = \frac{k}{t+1}, \qquad \tau = \ln \left( t + 1 \right).
\end{equation}
and 
\bel{nTex2}
n_k(t) = \frac{k^{k-1}}{k! \left(t + 1 \right)} \left(\frac{t}{t + 1}\right)^{k-1} \rme^{-kt/\left( t + 1 \right)}.
\ee
Again, we see much slower decay with time of the densities as compared to Smoluchowski solutions. In figure \ref{fig:tempdist}, we plot the distribution of temperatures of aggregates of different size at $t = 10$, obtained by Monte Carlo simulations and compare it with (\ref{eq:tempdist}). Again an excellent agreement between the numerical and exact analytical results confirm the accuracy of the exploited Monte Carlo method. 

Similar analysis may be performed for the model rate coefficients, which reduce the TDSE to the Smoluchowski equations with the size-multiplicative kernel. Generally, one can tailor  an infinite number of model rate kernels which allow exact solutions of TDSE. Some representative examples are given in the Table \ref{tab:exacttime}. The exact solutions presented in this table are essentially based on the exact solutions of standard Smoluchowski equations for the set of kernels, for which such solutions are available. Naturally, one can tailor another set of model rate coefficients based on other kernel for which exact solutions of Smoluchowski equations will be found. 

\begin{center}
\begin{table}[ht]
\caption{Some exact solutions with time-dependent temperatures for special cases of system (\ref{eq:gentempsys}).}
\label{tab:exacttime}
\begin{center}
\footnotesize
\begin{tabular}{|c|c|c|c|c|}
\hline
$C_{ij}$ \,\,\,& $B_{ij}$ \,\,\,& $D_{ij}$ \,\,\,& $T_k$ \,\,\,& $n_k$ \,\,\, \\
\hline
$T_i + T_j$ \,\,\,& $\left( T_i + T_j \right)^2/2$ \,\,\,& $\left( T_i + T_j \right)^2/2 \pm jT_i$ \,\,\,& $\rme^{\mp t}$ \,\,\,& \scriptsize $\left( \frac{ \pm (1 - \rme^{\mp t})}{1 \pm (1 - \rme^{\mp t})} \right)^{k-1} \left(1 \pm (1 - \rme^{\mp t})\right)^{-2}$ \,\,\,\\
\hline
$T_i + T_j$ & $\left( T_i + T_j \right)^2$ & $\left( T_i + T_j + 1 \right) T_i$ & $\frac{k}{1 + t}$ & $\frac{k^{k-1}}{k! \left(1 + t \right)} \left(\frac{t}{t + 1}\right)^{k-1} \rme^{-\frac{t}{t+1}}$ \\
\hline
$T_i T_j$ & $T_i T_j \left( T_i + T_j \right)$ & $T_i T_j \left( T_i + 1 \right)$ & $\frac{k}{1 + t}$ & $\frac{k^{k-3}}{\left( k - 1 \right)!} \left( \frac{t}{1+t} \right)^{k-1} \rme^{-kt/\left( 1 + t \right)}$ \\
\hline
$\frac{T_i}{i} + \frac{T_j}{j}$ & $\left( \frac{T_i}{i} + \frac{T_j}{j} \right) \left( T_i + T_j \right)$ & $\left( 2\frac{T_i}{i} + \frac{T_j}{j} \right) T_i$ & $\frac{k}{\sqrt{1 + 2t}}$ & $\left( 1 - \frac{1}{\sqrt{1 + 2t}} \right)^{k-1} \left(1 + 2t\right)^{-1}$ \\
\hline
$\frac{T_i}{i} + \frac{T_j}{j}$ & $\left( \frac{T_i}{i} + \frac{T_j}{j} \right) \left( T_i + T_j \right)$ & $\left( \frac{T_i}{i} + \frac{T_j}{j} - T_j \right) T_i$ & $\frac{k}{1 - t}$ & \scriptsize $\left( \frac{- \ln \left( 1 - t \right)}{1 - \ln \left( 1 - t \right)} \right)^{k-1} \left(1 - \ln \left( 1 - t \right)\right)^{-2}$ \\
\hline
\end{tabular}
\end{center}
\end{table}
\end{center}

\section{Conclusion}
We report exact solutions for the temperature-dependent Smoluchowski equations (TDSE), which describe the kinetics of ballistic agglomeration, where the evolution of densities of clusters of different size is entangled with the evolution of the partial temperatures (mean kinetic energy) of these clusters. Such solutions are important for the two reasons: Firstly, they may serve as the benchmark to assess the accuracy and computational efficiency of numerical schemes developed to solve TDSE. Secondly, the exact solution unambiguously demonstrate the variety of different evolution scenarios for the ballistic agglomeration, as the numerical solutions describe the behavior of the system only for a limited time interval. We analyze the evolution of the systems  with time-dependent temperatures for two general cases -- when all aggregates possess the same temperature and when temperatures of different species are different. In both cases, we exploit the exact solution of the common Smoluchowski equation and propose the model rate coefficients, that allow exact solutions of the TDSE. We report a wide variety of the evolution scenarios that demonstrate the obtained exact solutions: (i) permanent aggregation with permanent cooling; (ii) aggregative evolution with cooling to a jammed state; (iii) heating to infinite temperature during a final time, which resembles gelation, but with respect to kinetic energy; (iv) permanent aggregation with cooling down, or heating up  to a certain constant temperature; (v) gelation with cooling during a finite time. Note that the regimes of infinite increase of temperature and gelation may not be physical, as it has been demonstrated that the ballistic agglomeration with microscopically motivated rate kernels does not undergo gelation. In the paper, we present the derivation of the exact solutions for some most prominent model kernel and provide a couple of other exact solutions without a derivation. We compare our exact analytical solutions with the numerical solutions of the TDSE found by the Monte Carlo method and observe an excellent agreement between the numerical and exact results. In this way be confirm the accuracy of the new  Monte Carlo  approach. We believe that our results will help to better understand the nature of the ballistic agglomeration and will be used to check the accuracy of the respective numerical approaches, developed to solve the TDSE. 

\ack
The study was supported by a grant from the Russian Science Foundation No.~21-11-00363, https://rscf.ru/project/21-11-00363/.

\section*{Appendix A}
Here we present for completeness the microscopic expressions for the rate coefficients of temperature-dependent Smoluchowski equations. The detailed derivation and discussion of the physical meaning of these coefficients may be found in \cite{BFP2018,BOK_PRE20}. The coefficients $C_{ij}$, $B_{ij}$ and $D_{ij}$ read, where $\theta_i =T_i/m_i$ ($m_i=m_1 i$ is the mass of aggregate of size $i$):

\begin{eqnarray}
  {C_{ij}} = 2\sqrt {2\pi } \sigma _{ij}^2\sqrt {{\theta _i} + {\theta _j}} \left( {1 - {f_{ij}}} \right), \nonumber \\
  B_{ij} = 2\sqrt {2\pi} \sigma_{ij}^2\frac{m_i + m_j}{\sqrt {\theta _i + \theta _j} } \left( \theta _i\theta _j \left( 1 - f_{ij} \right) \right. \\
  \left. + \frac{4}{3}\left( \frac{i\theta_i - j\theta_j}{i + j} \right)^2 \left( 1 - g_{ij} \right) \right), \label{sys-coef} \\
  {D_{ij}} = 2\sqrt {2\pi } \sigma_{ij}^2\frac{m_i}{{\sqrt {{\theta _i} + {\theta _j}} }}\left( {{\theta _i}{\theta _j}\left( {1 - {f_{ij}}} \right)} \right. \\
  \left. + {\frac{4}{3}\theta_i^2\left( {1 - {g_{ij}}} \right)} \right. \\
  \left. { + \frac{4\left( 1 + \varepsilon \right) j}{3 \left( i + j \right)} \left( \theta_i + \theta_j \right) \left( {{\theta _i} - \frac{\left( 1 + \varepsilon \right) j}{2 \left( i + j \right)} \left( {{\theta _i} + {\theta _j}} \right)} \right)} g_{ij} \right), \nonumber
\end{eqnarray}
where 
\begin{eqnarray}
  {\sigma _{ij}} & = r_1\left( i^{1/d} + j^{1/d} \right), \nonumber \\
  {f_{ij}} & = \int\limits_{q_{ij}}^{\infty} c \rme^{-c} \, \rmd c = {\rme^{ - {q_{ij}}}}\left( {1 + {q_{ij}}} \right), \label{sys-coef2} \\
  {g_{ij}} & = \frac12 \int\limits_{q_{ij}}^{\infty} c^2 \rme^{-c} \, \rmd c = \rme^{ - q_{ij}}\left( 1 + q_{ij} + q_{ij}^2/2 \right), \nonumber
\end{eqnarray}
with $\sigma_{ij}$ being the collision cross-section (recall that the radius of a cluster of size $i$ reads $R_i=r_1i^{1/d}$, where $d$ is the cluster dimension), 
and $q_{ij}$ characterizes the average  ratio of potential and relative kinetic energy: 
\be\label{qij}
{q_{ij}}  = \frac{{{W_{ij}}}}{{{\varepsilon ^2} \mu_{ij} \left( {{\theta _i} + {\theta _j}} \right)}} \ee
where, $\mu_{ij}= m_i m_j /(m_i+m_j) =m_1ij/(i+j)$ is the reduced mass of a colliding pair ($m_1$ is the monomer mass), and $\varepsilon$ is the restitution coefficient. $W_{ij}$ in (\ref{qij}) describes the interaction energy barrier for two aggregates of size $i$ and $j$ at a contact:
\be\label{W}
W \left(i, j \right) = W_{ij} = a\frac{{{{\left( {{i^{1/3}}{j^{1/3}}} \right)}^{{\lambda _1}}}}}{{{{\left( {{i^{1/3}} + {j^{1/3}}} \right)}^{{\lambda _2}}}}}.
\ee 
Here the constant $a$ specifies the interaction energy, while $\lambda_{1} $ and $\lambda_{2}$ quantify the dependence of $W_{ij}$ on the size of particles $i$ and $j$. For instance, $\lambda_1=\lambda_2=4/3$ correspond to the adhesive surface interactions,  $\lambda_1=\lambda_2=3$ stands for the dipole-dipole interactions  and $\lambda_1=3$, $\lambda_2=1$ refers to the gravitational or Coulomb interaction, when the particles charges scale as their masses \cite{BFP2018}.

For the case of energy equipartition of different species, $T_i=T$ for all $i$, the rate coefficients $P_{ij}$  read, 
$$
P_{ij}=\frac12\left[D_{ij}+D_{ji} - B_{ij} \right].
$$
If the energy barrier is large, as compared with the temperature $T$, that is, $W_{ij} \gg T$, then (\ref{sys-coef}) -(\ref{W}) yield much simpler rate coefficients   \cite{BFP2018}:
\beel{CT}
C_{ij} &=& 2 \sigma_{ij}^2 \sqrt{2 \pi T/\mu_{ij} } \\
P_{ij} &=& \frac23 T \, C_{ij}. 
\eeq

\section*{Appendix B}

Here we give a brief description of the temperature-dependent Monte Carlo method. In short, it does the same steps as the Monte Carlo for classic Smoluchowski equations, see e.g. \cite{MCSmol,Babovsky,MCSmol2,kalinov2021direct}, except one more step, where the partial temperatures are updated.

Let a system of volume $V$ contain $N_i$ particles of size $i$, so that $n_i = N_i / V$. Let $T_i$ be the temperatures of size-$i$ particles. Collisions in the temperature-dependent Monte Carlo method are performed as follows:
\begin{enumerate}
    \item Choose the pair of particles $(i,j)$ with the probability $p_{ij} = \frac{C_{ij} N_i N_j}{\sum\limits_{k,l} C_{kl} N_k N_l}$, where $C_{ij}$ are the according rate constants.
    \item Advance the time as $t := t + \Delta t$, so that the average time between collisions be  $\left\langle \Delta t \right\rangle = \frac{V}{\frac{1}{2} \sum\limits_{k,l} C_{kl} N_k N_l}$.
    \item Update the temperatures $T_i$, $T_j$, $T_{i+j}$ as explained below.
    \item Update the particle numbers due to aggregation:
    \[
       N_i := N_i - 1, \quad N_j := N_j - 1, \quad N_{i+j} := N_{i+j} + 1.
    \]
    \item Replicate  all particles, when the total number of particles halves.
\end{enumerate}

Apart from the temperature updates, all other steps are the same as for classical Smoluchowski equations. The choice of the colliding particles according to the probabilities $p_{ij} \sim C_{ij} N_i N_j$ can be accomplished using the standard acceptance-rejection technique \cite{monte-random}. Or, to make it faster, one can use the low-rank approach, described in \cite{OB_PRE2022}.

The only difference is the third step, where we update the temperatures, which then affect the collision rates $C_{ij} = C_{ij} (T_i, T_j)$ for further collisions. We use the following updates \cite{OB_PRE2022}:
\begin{eqnarray}
  T_i & := \frac{N_i T_i - D_{ij} / C_{ij}}{N_i - 1}, \label{eq:tiupd} \\
  T_j & := \frac{N_j T_j - D_{ji} / C_{ji}}{N_j - 1}, \label{eq:tjupd} \\
  T_{i+j} & := \frac{N_{i+j} T_i + B_{ij} / C_{ij}}{N_{i+j} + 1}.
\end{eqnarray}

Note that we always solve the original TDSE, even when studying the simplified models with temperature equipartition. Thus, we do not make temperature equi\-partition assumption in the Monte Carlo simulations and obtain it naturally, when it happens (up to the usual stochastic noise, common in Monte Carlo simulations).

One of the main advantages of the temperature-dependent Monte Carlo method is that the stochastic noise comes only from the number density discretization. The temperature updates, on the other hand, depend only on the sizes of the colliding particles, while in reality particles of the same size can have different speeds, so the changes of the partial temperatures (average kinetic energies) also depend on whether the speeds of the colliding particles are lower or higher than the average.

Some disadvantages of this method are related to the possibility, that the updates (\ref{eq:tiupd}-\ref{eq:tjupd}) can, in principle, lead to negative temperatures (they should be then rounded up to zero). This occurs very rarely, unless one deliberately explores the case, when all temperatures quickly drop to zero, as it happens in the jammed state. In this case the rounding errors add up. That's why we have been forced to use $10^5$ particles for the system shown in Fig. \ref{fig:jam}, while $10^4$ particles was already enough to get a good agreement with the exact solutions for all other cases.

\newpage

\bibliography{exactsolTSm}
\end{document}